\documentclass{journalA4}
\usepackage{latexsym,graphicx,wrapfig,fancyhdr,url,subfigure,times,mathptm}

\DeclareSymbolFontAlphabet{\Bbb}{AMSb}
\def\R{\ensuremath{\Bbb R}}

\newcommand{\graph}{\mathcal{G}}
\newcommand{\layout}{\mathcal{L}}

\newcommand{\tree}{\mathcal{T}}

\graphicspath{{figures/}}

\title{Area-Universal Rectangular Layouts}

\author{David Eppstein$^1$ \and Elena Mumford$^2$ \and Bettina Speckmann$^2$ \and Kevin Verbeek$^2$}

\date{\small $^1$ Department of Computer Science, University of California, Irvine, USA.\\{\tt eppstein@ics.uci.edu}\\
$^2$ Department of Mathematics and Computer Science, TU
Eindhoven, The Netherlands.\\ {\tt e.mumford@tue.nl}, {\tt speckman@win.tue.nl}, and {\tt k.a.b.verbeek@tue.nl}}

\begin{document}
\maketitle

\begin{abstract}\noindent
A rectangular layout is a partition of a rectangle into a finite set of interior-disjoint rectangles. Rectangular layouts appear in various applications: as rectangular cartograms in cartography, as floorplans in building architecture and VLSI design, and as graph drawings. Often areas are associated with the rectangles of a rectangular layout and it might hence be desirable if one rectangular layout can represent several area assignments. A layout is \emph{area-universal} if any assignment of areas to rectangles can be realized by a combinatorially equivalent rectangular layout. We identify a simple necessary and sufficient condition for a rectangular layout to be area-universal: a rectangular layout is area-universal if and only if it is \emph{one-sided}. More generally, given any rectangular layout $\layout$ and any assignment of areas to its regions, we show that there can be at most one layout (up to horizontal and vertical scaling) which is combinatorially equivalent to $\layout$ and achieves a given area assignment.
We also investigate similar questions for perimeter assignments. The adjacency requirements for the rectangles of a rectangular layout can be specified in various ways, most commonly via the dual graph of the layout. We show how to find an area-universal layout for a given set of adjacency requirements whenever such a layout exists.
\end{abstract}

\section{Introduction}

{\bf Motivation.} Raisz~\cite{Rai-GR-34} introduced \emph{rectangular cartograms} in 1934 as a way of visualizing spatial information, such as  population or economic strength, of a set of regions like countries or states. Rectangular cartograms represent geographic regions by rectangles; the positioning and adjacencies of these rectangles are chosen to suggest their geographic locations to the viewer, while their areas are chosen to represent the numeric values being communicated by the cartogram. The stylization inherent in replacing the complicated shapes of geographic regions by rectangles is a feature of such diagrams: as Raisz writes, ``simple distortion of the map would be misleading,'' because it is important to emphasize that a cartogram is not a map.

Often more than one numeric quantity should be displayed as a cartogram for the same set of geographic regions. The first three figures Raisz shows, for instance, are cartograms of land area, population, and wealth within the United States. To make the visual comparison of multiple related cartograms easier, it is desirable that the arrangement of rectangles be combinatorially equivalent in each cartogram, although the relative sizes of the rectangles will differ. This naturally raises the question: when is this possible?

\begin{figure}[b]
\centering\includegraphics{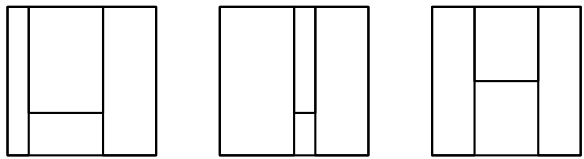}
\caption{Three area assignments for an area-universal layout.}
\label{fig:area-universal}
\end{figure}
Mathematically, a rectangular cartogram is a \emph{rectangular layout}: a partition of a rectangle into finitely many interior-disjoint rectangles. We call a layout $\layout$ \emph{area-universal} if, no matter what areas we require each of its regions to have, some combinatorially equivalent layout $\layout'$ has regions with the specified areas. For instance, the four-region rectangular layout shown below
with three different area assignments is area-universal: any four numbers can be used as the areas of the rectangles in a combinatorially equivalent layout.

Area-universal rectangular layouts are useful not only for displaying multiple side-by-side cartograms for different sets of data on the same regions, but also for dynamically morphing from one cartogram into another. Additionally, rectangular layouts have other applications in which being able to choose a layout first and then later assigning varying areas while keeping the combinatorial type of the layout fixed may be an advantage: in circuit layout applications of rectangular layouts~\cite{YeaSar-SJDM-95}, each component of a circuit may have differing implementations with differing tradeoffs between area, energy use, and speed; in building design it is desirable to be able to determine the areas of different rooms according to their function~\cite{Earl1979}; in treemap visualizations, alternative area-universal layouts may be of use in controlling rectangle aspect ratios~\cite{BruHuiWij-DV-00}; and in graph drawing applications~\cite{KanHe-TCS-97} the areas of rectangles may need to vary according to the labels or other features to be placed in the drawing. Thus, it is of interest to identify the properties that make a rectangular layout area-universal, and to find area-universal layouts when they exist.

\begin{wrapfigure}[10]{r}{.4\textwidth}
  \centering
  \includegraphics{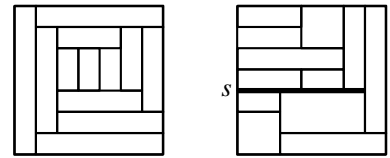}
  \caption{The left layout is one-sided, but the right one is not: the maximal segment $s$ is not the side of any rectangle.}
  \label{fig:one-sided}
\end{wrapfigure}
\smallskip\noindent
{\bf Results.} We identify a simple necessary and sufficient condition for a rectangular layout to be area-universal: a rectangular layout is area-universal if and only if it is \emph{one-sided}. One-sided layouts are characterized via their \emph{maximal line segments}. A line segment of a layout $\layout$ is formed by a sequence of consecutive inner edges of $\layout$. A segment of $\layout$ that is not contained in any other segment is maximal. In a one-sided layout every maximal line segment $s$ must be the side of at least one rectangle $R$; any vertices interior to $s$ are T-junctions that all have the same orientation, pointing away from $R$ (Figure~\ref{fig:one-sided}).
Given an area-universal layout $\layout$ and an assignment of areas for its regions, we describe a numerical algorithm that finds a combinatorially equivalent layout $\layout'$ whose regions have a close approximation to the specified areas. These results can be found in Section~\ref{sec:onesided}.

More generally, given any rectangular layout $\layout$ and any assignment of areas to its regions, we show in Section~\ref{sec:onlyone} that there can be at most one layout (up to horizontal and vertical scaling) which is combinatorially equivalent to $\layout$ and achieves the given area assignment. This result was previously known only for two special classes of rectangular layouts, namely \emph{sliceable layouts} (layouts that can be obtained by recursively partitioning a rectangle by horizontal and vertical lines) and \emph{L-shape destructable layouts}~\cite{KreSpe-CGTA-07} (layouts where the rectangles can be iteratively removed such that the remaining rectangles form an L-shaped polygon).

In Section~\ref{sec:perimeter} we investigate \emph{perimeter cartograms} in which the perimeter of each rectangle is specified rather than its area. Again, any rectangular layout can have at most one combinatorially equivalent layout for a given perimeter assignment; it is possible in polynomial time to find this equivalent layout, if it exists.

\begin{wrapfigure}[11]{r}{.35\textwidth}
\centering
\includegraphics{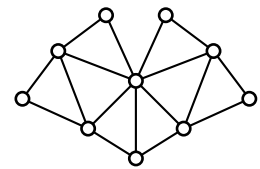}
\vspace{-.25\baselineskip}
\caption{A graph that is not the dual of an area-universal layout: the rectangle dual to the bottom center vertex may not be arbitrarily large~\cite{Rin-EPB-87}.}
\label{fig:rinsma-graph}
\end{wrapfigure}
The rectangles of a rectangular cartogram should have the same adjacencies as the regions of the underlying map. Hence, the dual graph of the cartogram should be the same as the dual graph of the map. Here, the \emph{dual graph} is the graph that has one node per region and connects two regions if they are adjacent, where two regions are considered to be adjacent if they share a 1-dimensional part of their boundaries. The dual of a rectangular cartogram or layout must be a triangulated plane graph satisfying certain additional conditions. We call such graphs \emph{proper graphs} (see Section~\ref{sec:def} for a detailed definition). Every proper graph $\graph$ has at least one \emph{rectangular dual}: a rectangular layout $\layout$ whose dual graph is $\graph$. However, not every proper graph has an area-universal rectangular dual; Rinsma~\cite{Rin-EPB-87} described an outerplanar proper graph $\graph$ and an assignment of weights to the vertices of $\graph$ such that no rectangular dual of $\graph$ can have these weights as the areas of its regions (Figure~\ref{fig:rinsma-graph}). Thus, it is of interest to determine which proper graphs have an area-universal rectangular dual.  In Section~\ref{sec:Findonesided} we describe algorithms that, given a proper graph $\graph$, find an area-universal rectangular dual of $\graph$ if it exists. These algorithms are not fully polynomial, but are fixed-parameter tractable for a parameter related to the number of separating four-cycles in $\graph$.

Motivated by architectural plans, where only a subset of the room adjacencies might be specified, Ri\-nsma~\cite{Rinsma1988} considered a weaker version of the problem described above: given a tree $\tree$ does there exist a rectangular layout $\layout$ such that $\tree$ is a spanning tree of the dual graph of $\layout$? She showed that such a layout always exists, but the layouts constructed by her algorithm are not necessarily area-universal. In Section~\ref{sec:dualtree} we modify her construction to yield area-universal layouts, proving that for every tree $\tree$ there is an area-universal layout $\layout$ such that $\tree$ is a spanning tree of the dual graph of $\layout$.

\section{Preliminaries}\label{sec:def}

As stated above, a rectangular layout (or sometimes simply \emph{layout}) is a partition of a rectangle into a finite set of interior-disjoint rectangles. We assume that no four regions meet in a single point, as it is true (with a notable exception in the American Southwest) for most geographic partitions of interest.
We denote the dual graph of a layout $\layout$ by $\graph(\layout)$. A layout $\layout$ such that $\graph = \graph(\layout)$ is called a \emph{rectangular dual} of graph $\graph$. $\graph(\layout)$ is a plane triangulated graph and is unique for any layout $\layout$. Not every plane triangulated graph has a rectangular dual, and if it does, then the rectangular dual is not necessarily unique.

\begin{figure}[h]
  \centering
  \includegraphics{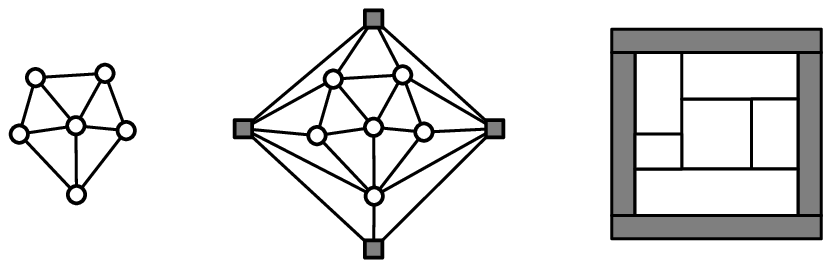}
  \caption{A proper graph $\graph$, an extended graph $E(\graph)$, and a rectangular dual $\layout$ of $E(\graph)$.}
\label{fig:rect-dual}
\end{figure}

Kozminski and Kinnen~\cite{KozKin-Nw-85} proved that a plane triangulated graph $\graph$ has a rectangular dual if and only if we can augment $\graph$ with four external vertices in such a way that the \emph{extended graph} $E(\graph)$ has the following two properties: $(i)$ every interior face is a triangle and the exterior face is a quadrangle; $(ii)$ $E(\graph)$ has no separating triangles---a \emph{separating triangle} is a separating cycle (a simple cycle that has vertices both inside and outside) of length three.\footnote{More generally, we call a  separating cycle of length $k$ a \emph{separating $k$-cycle}.} If a plane triangulated graph $\graph$ allows such an augmentation, then we say that $\graph$ is a \emph{proper graph}. A rectangular dual of an extended graph of a proper graph $\graph$ can be constructed in linear time~\cite{KanHe-TCS-97} and it immediately implies a rectangular dual for $\graph$ (Figure~\ref{fig:rect-dual}).

An extended graph $E(\graph)$ determines uniquely which vertices of a proper graph $\graph$ are associated with the corner rectangles of every rectangular dual of $\graph$ that corresponds to $E(\graph)$. For a given proper graph there might be several possible extended graphs and hence several possible \emph{corner assignments}. In many cases we assume that a corner assignment, and hence an extended graph, has already been fixed, but if this is not the case then it is possible to test all corner assignments, as there can be only polynomially many of them.

\begin{wrapfigure}[11]{r}{.39\textwidth}
\vspace{-.5\baselineskip}
\centering\includegraphics{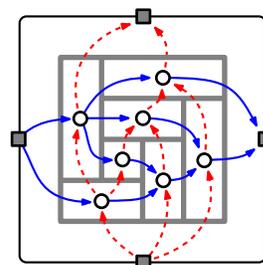}
\vspace{-.35\baselineskip}
\caption{A rectangular layout and the regular edge labeling of its extended dual.}
\label{fig:REL}
\end{wrapfigure}
A rectangular layout $\layout$ naturally induces a labeling of its extended dual graph $E(\graph)$. If two rectangles of $\layout$ share a vertical segment, then we color the corresponding edge in $E(\graph)$ blue (solid) and direct it from left to right. Correspondingly, if two rectangles of $\layout$ share a horizontal segment, then we color the corresponding edge in $E(\graph)$ red (dashed) and direct it from bottom to top (Figure~\ref{fig:REL}).

This labeling has the following properties: $(i)$ around each inner vertex in clockwise order we have four contiguous sets of incoming blue edges, outgoing red edges, outgoing blue edges, and incoming red edges; $(ii)$ the left exterior vertex has only blue outgoing edges, the top exterior vertex has only red incoming edges, the right exterior vertex has only blue incoming edges, and the bottom exterior vertex has only red outgoing edges.

Such a labeling is called a \emph{regular edge labeling}. It was introduced by Kant and He~\cite{KanHe-TCS-97} who showed that every regular edge labeling of an extended graph $E(\graph)$ uniquely defines an equivalence class of rectangular duals of a proper graph $\graph$. Given any extended graph $E(\graph)$, a regular edge labeling for $E(\graph)$ can be found in linear time and the rectangular dual defined by it can also be constructed in linear time~\cite{KanHe-TCS-97}. Regular edge labelings have also been studied by Fusy~\cite{Fus-GD-05,Fus-DM-08}, who refers to them as \emph{transversal structures}.

Two layouts $\layout$ and $\layout'$ are \emph{equivalent}, denoted by $\layout \sim \layout'$, if they  induce the same regular edge labeling of the same dual graph. We say that a rectangular layout $\layout$ with $n$ rectangles $R_1,..., R_n$ realizes a weight function $w: {R_1,...,R_n} \rightarrow \R, w(i) > 0$ as a \emph{rectangular cartogram} if there exists a layout $\layout' \sim \layout$ such that for any $1 \leq i \leq n$ the area of rectangle $R_i$ equals $w(r_i)$. Correspondingly, we say that a layout $\layout$ realizes $w$ as a \emph{perimeter cartogram} if there exists a layout $\layout' \sim \layout$ such that the perimeter of each rectangle of $\layout'$ equals the prescribed weight. A layout $\layout$ is \emph{area-universal} if it realizes every possible weight function.

\begin{wrapfigure}[7]{r}{.39\textwidth}
\centering\includegraphics{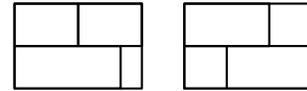}
\vspace{-.25\baselineskip}
\caption{Two inequivalent but order-equivalent rectangular layouts.}
\label{fig:ordereqv}
\end{wrapfigure}
It will be convenient to define a weaker equivalence relation on layouts than equivalence, which we call \emph{order-equivalence}. For a layout $\layout$, we define a partial order on the vertical maximal segments, in which $s_1\le s_2$ if there exists an $x$-monotone curve that has its left endpoint on $s_1$, its right endpoint on $s_2$, and that does not cross any horizontal maximal segments. This partial order can equivalently be defined by a directed acyclic multigraph that has a vertex per maximal segment and an edge from the segment on the left boundary of each rectangle to the segment on the right boundary of the same segment; this graph is an $st$-planar graph, a planar DAG in which the unique source and the unique sink are both on the outer face. The dual of this $st$-planar graph defines in a symmetric way a partial order on the horizontal maximal segments. We say that $\layout$ and $\layout'$ are order-equivalent if their rectangles and maximal segments correspond one-for-one in a way that preserves these partial orders.

\begin{obs}\label{lem:dof}
A rectangular layout with $n$ rectangular regions has $n-1$ maximal segments.
\end{obs}

\section{There can be only one}\label{sec:onlyone}

We first show that for any combination of layout and weight function there can be at most one rectangular cartogram or perimeter cartogram. More generally, if two geometrically different but order-equivalent layouts share the same bounding box, there is a rectangle in one of the layouts that is larger in both of its dimensions than the corresponding rectangle in the other layout. The proof involves a graph-theoretic argument in an auxiliary graph constructed from the two layouts.

\begin{figure}[h]
\centering\includegraphics{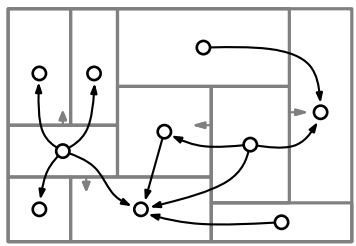}
\caption{A push graph. The layout $\layout$ is shown; the relative position of the maximal segments in the equivalent layout $\layout'$ is indicated by the arrows attached to the maximal segments.}
\label{fig:pushgraph}
\end{figure}

Thus, let $\layout$ and $\layout'$ be two geometrically different order-equivalent layouts with the same bounding box. The \emph{push graph} $\mathcal{H}$ of $\layout$ and $\layout'$ is a directed graph that has a vertex for each rectangle in $\layout$ and an edge from vertex $R_i$ to vertex $R_j$ if the rectangles $R_i$ and $R_j$ are adjacent and the maximal segment in $\layout$ that separates $R_i$ from $R_j$ is shifted in $\layout'$ towards $R_j$ and away from $R_i$ (Figure~\ref{fig:pushgraph}) .

\begin{figure}[t]
\centering\includegraphics{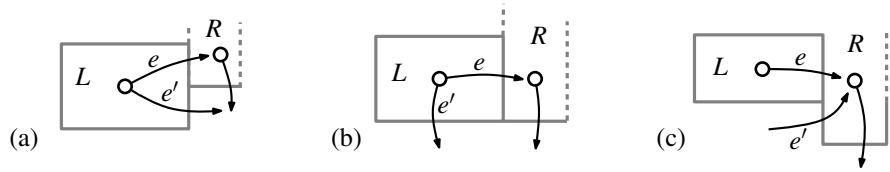}
\caption{Cases for Lemma~\ref{lem:push}.}
\label{fig:pushcases}
\end{figure}
\begin{lemma}
\label{lem:push}
The push graph for $\layout$ and $\layout'$ contains a node with no incoming or no outgoing edges.
\end{lemma}
\begin{proof}
Assume for contradiction that the push graph $\mathcal{H}$ has no source or sink. Then $\mathcal{H}$ must contain a cycle. Let $C$ be a simple cycle in $\mathcal{H}$ that encloses as few vertices as possible, and assume without loss of generality that $C$ is oriented clockwise. By construction, $C$ cannot contain a rightward edge immediately followed by a leftward edge or an upward edge immediately followed by an downward edge. Hence it must contain a rightward edge $e$ that is followed by a downward edge.
We distinguish three cases depending on the relative positions of the bottom sides of the two rectangles $L$ and $R$ that are connected by $e$ (Figure~\ref{fig:pushcases}):
\begin{itemize}\topsep=1pt\itemsep=0pt\parsep=0pt
\item[(a)] If the bottom edge of $L$ lies below the bottom edge of $R$, then $\mathcal{H}$ must contain an edge $e'$ that connects $L$ to the rectangle below $R$. This edge $e'$ shortcuts $C$, contradicting the minimality of $C$.
\item[(b)] If the bottom edges of $L$ and $R$ are aligned along a maximal segment, then $\mathcal{H}$ must contain an edge $e'$ that points downward from $L$. By following a directed chain of edges starting with $e'$ we either reach a repeated vertex within this chain of edges, or a vertex that belongs to $C$. In either case we have found a cycle that encloses fewer vertices than $C$, contradicting the minimality of $C$.
\item[(c)] If the bottom edge of $L$ lies above the bottom edge of $R$, then $\mathcal{H}$ must contain an edge $e'$ that connects the rectangle below $L$ to $R$. As in case (b) by following a chain of edges backwards starting from $e'$ we can find a cycle that encloses fewer vertices than $C$, contradicting the minimality of $C$.
\end{itemize}
\end{proof}

\begin{theorem}
\label{thm:area-uniqueness}
For any layout $\layout$ and any weight function $w$ there is at most one layout $\layout'$ (up to affine transformations) that is order-equivalent to $\layout$ and that realizes $w$ as a rectangular cartogram.
\end{theorem}
\begin{proof}
Let $\layout$ and $\layout'$ be order-equivalent with the same area, but geometrically different; scale $\layout'$ horizontally and vertically so they have the same bounding box.
Lemma~\ref{lem:push} implies that one of the layouts contains a rectangle $R$ that is at least as large both horizontally and vertically, and strictly larger in one of the two dimensions, than the corresponding rectangle of the other. Thus, $R$ cannot have the same area in both layouts and only one of the layouts can realize~$w$.
\end{proof}
\begin{wrapfigure}[8]{r}{.35\textwidth}
\centering\vspace{-.75\baselineskip}
\includegraphics{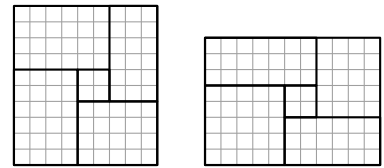}
\vspace{-.25\baselineskip}
\caption{Two equivalent layouts in which corresponding rectangles have the same perimeter.}
\vspace{-.75\baselineskip}
\label{fig:nuperim}
\end{wrapfigure}

For perimeter, such strong uniqueness does not hold: there are equivalent layouts that are not affine transformations of each other in which the perimeters of corresponding rectangles are equal (Figure~\ref{fig:nuperim}). However, if we fix the outer bounding box of the layout, the same proof method works:%
\begin{theorem}
\label{thm:perim-uniqueness}
For any layout $\layout$ and any weight function $w$ there is at most one layout $\layout'$ that is order-equivalent to $\layout$ with the same bounding box and that realizes $w$ as a perimeter cartogram.
\end{theorem}
More generally the same result holds for any type of cartogram in which rectangle sizes are measured by any strictly monotonic function of the height and width of the rectangles.

\section{Area-universality and one-sidedness}\label{sec:onesided}
As the next lemma states, all layouts are area-universal in a weak sense involving order-equivalence in place of equivalence. The proof uses Theorem~\ref{thm:area-uniqueness} to invert the map from vectors of positions of segments in a layout to vectors of rectangle areas, along a line segment from the area vector of $\layout$ to the desired area vector.
\begin{lemma}
\label{lem:existence}
For any layout $\layout$ and weight function $w$, there exists a layout $\layout'$ that has a square outer rectangle, is order-equivalent to $\layout$, and realizes $w$ as a rectangular cartogram.
\end{lemma}
\begin{proof}
The outer rectangle of $\layout'$ is uniquely determined (up to congruence) by having a square shape and having an area equal to the sum of the weights. By scaling horizontally and vertically, we may assume without loss of generality that $\layout$ has the same outer square; let $s$ be the side length of this square, and without loss of generality let it be placed in a Cartesian plane with the coordinates $\{(x,y)\mid 0\le x,y\le s\}$.

Coordinatize the space of layouts that are order-equivalent to $\layout$ with the same bounding box by supplying a Cartesian coordinate $c_i$ for each maximal segment of $\layout$: its $x$-coordinate for a vertical maximal segment, or its $y$-coordinate for a horizontal maximal segment. These coordinates satisfy the linear constraints $0<c_i<s$; additionally, each rectangle having segment $i$ on its left side and segment $j$ on its right side (or segment $i$ on its bottom side and segment $j$ on its top side) corresponds to a constraint $c_i<c_j$. Conversely, any assignment of coordinates $c_i$ satisfying these constraints determines a layout that is order-equivalent to $\layout$. This finite set of linear inequalities is satisfied by the points in an open convex polytope $P$; by Observation~\ref{lem:dof}, $P$ has dimension $n-1$ where $n$ is the number of rectangles in the layout.

Consider the quadratic function $W$ that maps a point $p$ in $P$ to the weight function describing the areas of the rectangles in the layout corresponding to $p$. The domain and range of this function are both $(n-1)$-dimensional: there are $n$ rectangle areas to determine, but the total area is fixed, so the image of $W$ lies in an $(n-1)$-dimensional linear subspace of $\R^n$.  By Theorem~\ref{thm:area-uniqueness} this function $W$ is one-to-one in $P$. Because it is just a quadratic function, $W$ can be extended to all of $\R^{n-1}$ and in particular to the closure of $P$; the points on the boundary of $P$ are mapped by $W$ to improper weight functions in which the weight of some rectangle is zero. We need to show  that $w$ is in the image of $P$. We show more generally that the whole line segment $W(\layout)w$ is within the image of $P$.  Let $w'$ be the farthest point from $W(\layout)$ on this line segment such that the open line segment $W(\layout)w'$ is within the image of $P$. The inverse image of segment $W(\layout)w'$ forms a curve within $P$ that has as its endpoints $\layout$ and some other layout $\layout^*$ such that $W(\layout^*)=w'$. $\layout^*$ must be interior to $P$ because
$w'$ is a linear interpolant of $W(\layout)$ and $w$ and therefore has all weights nonzero. Thus, $w'$ itself is also within the image of $P$. If $w=w'$ we are done. Otherwise, in a neighborhood of $w'$, $W$ is a smooth one-to-one function from and to an $(n-1)$-dimensional space, and hence is invertible; we may use this local inverse to extend the inverse image of segment $W(\layout)w'$ past $w'$, contradicting the assumption that $w'$ is the farthest point from $W(\layout)$ to which the inverse image can be extended and completing the proof.
\end{proof}

One may find $\layout'$ by hill-climbing to reduce the Euclidean distance between the current weight function and the desired weight function. No layout $\layout$ can be locally but not globally optimal, because within any neighborhood of $\layout$ the inverse image of the line segment connecting its weight vector to the desired weight vector contains layouts that are closer to $w$. Alternatively, one can find $\layout'$ by a numerical procedure that follows this inverse image by inverting the Jacobean matrix of $W$ at each step. We do not know whether it is always possible to find $\layout'$ exactly by an efficient combinatorial algorithm (as may easily be done for the subclass of sliceable layouts), or whether the general solution involves roots of high-degree polynomials that can be found only numerically.

\begin{theorem}
\label{thm:3-equivalent-properties}
The following three properties of a layout $\layout$ are equivalent:
\begin{enumerate}\itemsep0pt
\item $\layout$ is area-universal.
\item Every layout that is order-equivalent to $\layout$ is equivalent to $\layout$.
\item $\layout$ is one-sided.
\end{enumerate}
\end{theorem}

\begin{proof}
We show that property 2 implies property 1, that the negation of property 3 implies the negation of property 1, that property 3 implies property 2, and that the negation of property 3 implies the negation of property 2.

$2\Rightarrow 1$: Let $\layout$ be a layout satisfying the property that every layout that is order-equivalent to $\layout$ is equivalent to $\layout$, and let $w$ be an arbitrary weight function; we must show that $\layout$ realizes $w$ as a rectangular cartogram. By Lemma~\ref{lem:existence}, there exists a layout $\layout'$ that is order-equivalent to $\layout$ and realizes $w$; by the assumption, $\layout'$ is equivalent to $\layout$, as desired.

$(\lnot 2)\Rightarrow (\lnot 1)$: Suppose there exists a layout $\layout'$ that is order-equivalent but inequivalent to $\layout$. By scaling horizontally and vertically, we may assume that $\layout$ and $\layout'$ have the same bounding box. Let $w$ be the weight function given by the areas of the rectangles in $\layout'$. By Theorem~\ref{thm:area-uniqueness}, $\layout'$ is the only layout that is order-equivalent to $\layout$ and realizes $w$ as a rectangular cartogram; therefore, there can be no layout that is equivalent to $\layout$ and realizes $w$ as a rectangular cartogram, showing that $\layout$ is not area-universal.

$3\Rightarrow 2$: Let $\layout$ be a one-sided layout, and $\layout'$ be order-equivalent to $\layout$. Then $\layout'$ must be one-sided, because the property of each maximal segment being a side of a rectangle is preserved under order-equivalence. For every pair of adjacent rectangles $R_1$ and $R_2$ in $\layout$ or in $\layout'$,  $R_1$ and $R_2$ are adjacent with a given orientation if and only if they are on opposite sides of a common maximal segment with the given orientation, and this property of being on opposite sides of a common maximal segment is also preserved by order-equivalence, so order-equivalence preserves the adjacencies of rectangles in $\layout$ and $\layout'$.

$(\lnot 3)\Rightarrow (\lnot 2)$: If $\layout$ is not one-sided, let $s$ be a maximal segment of $\layout$ that has more than one rectangle on both sides of $s$; without loss of generality assume that $s$ is horizontal. We may form an order-equivalent but inequivalent layout $\layout'$ by moving the vertical maximal segments that abut the top side of $s$ rightwards and the vertical maximal segments that abut the bottom side of $s$ leftwards until the order of their endpoints changes, as in Figure~\ref{fig:ordereqv}.
\end{proof}

\section{Finding perimeter cartograms}\label{sec:perimeter}

\begin{figure}[h]
  \centering
  \includegraphics{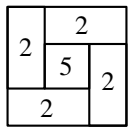}
\caption{The outer rectangles each contribute at most one unit of shared boundary to the perimeter of the central rectangle, which is too large to be realized.}
\label{fig:badperim}
\end{figure}

Although our proof of uniqueness for rectangular cartograms generalizes to perimeter, our
proof that any layout and weight function have a realization as an order-equivalent cartogram does not generalize: there exist one-sided layouts and weight functions that cannot be realized as a perimeter cartogram (Figure~\ref{fig:badperim}).

Nevertheless, one can test in polynomial time whether a solution exists for any layout
and weight function. The technique involves describing the constraints on the perimeters of rectangles as linear
equalities that reduce the dimension of the space of layouts to at most two, and forming a low-dimensional linear program from inequality constraints expressing the equivalence to $\layout$ of the other layouts within this low-dimensional  space.
\begin{theorem}
\label{thm:perim-alg}
For any layout $\layout$ and any weight function $w$ we can find a layout $\layout'$ that is equivalent to $\layout$ and that realizes $w$ as a perimeter cartogram, if one exists.
\end{theorem}
\begin{proof}
As in the proof of Lemma~\ref{lem:existence}, we may specify a layout by supplying one coordinate per maximal segment; together with the length and height of the bounding box this gives us a set of $n+1$ real values to be determined in a way consistent with the given weight function and layout. Each value of the weight function determines an equality constraint among these variables, stating that a certain linear combination of differences of segment positions equals the given perimeter. The constraints that the resulting layout be equivalent to $\layout$ may be translated into linear inequality constraints, stating that the segment on the left side of each rectangle must have a smaller coordinate value than the segment on the right, the segment on the bottom side of each rectangle must have a smaller coordinate value than the segment on the top, and that the three-way junctions appearing along any maximal segment of the layout appear in the correct order.

The equality constraints determine a linear subspace $S$ of $\R^{n+1}$ which we may find by Gaussian elimination. If there exists a layout $\layout'$ realizing $w$, then, by Theorem~\ref{thm:perim-uniqueness}, $S$ contains only a single point with the same bounding box height and width as $\layout'$, and hence has dimension at most two; conversely, if the dimension of this linear subspace is greater than two we may immediately infer from Theorem~\ref{thm:perim-uniqueness} that no solution exists.

If the dimension of the subspace is at most two, on the other hand, we may translate all the inequality constraints in $\R^{n+1}$ into linear inequality constraints in this two-dimensional subspace, and solve the resulting two-dimensional linear program in linear time using standard algorithms (e.g. see~\cite{Meg-JACM-84}).
\end{proof}

The same algorithm can be used to find an order-equivalent layout rather than an equivalent layout, by restricting the inequality constraints to the subset that determine order-equivalence.

\section{Finding one-sided layouts}\label{sec:Findonesided}

Recall that every proper triangulated plane graph has a rectangular dual, but not necessarily a one-sided rectangular dual. Since one-sided duals are area-universal, it is of interest to find a one-sided dual for a proper graph if one exists. Our overall approach is, first, to partition the graph on its separating four-cycles; second, to represent the family of all layouts for a proper graph as a distributive lattice, following Fusy~\cite{Fus-GD-05,Fus-DM-08}; third, to represent elements of the distributive lattice as partitions of a partial order according to Birkhoff's theorem~\cite{Bir-DMJ-37}; fourth, to characterize the ordered partitions that correspond to one-sided layouts; and fifth, to search in the partial order for partitions of this type.
Our algorithms are not fully polynomial, but they are polynomial whenever the number of separating four-cycles in the given proper graph is bounded by a fixed constant, or more generally when such a bound can be given separately within each of the pieces found in the partition we find in the first stage of our algorithms.

\subsection{Eliminating nontrivial separating four-cycles}

Recall that a \emph{separating four-cycle} in a plane graph $\graph$ is a cycle of four vertices that has other vertices both inside and outside it. We say that a separating four-cycle is \emph{nontrivial} if the number of vertices inside it is greater than one. Although a plane graph may have a quadratic number of separating four-cycles (for instance this is true for the complete bipartite graph $K_{2,n-2}$) it is possible to represent all separating four-cycles in linear space by finding all maximal complete bipartite subgraphs $K_{2,i}$ of $\graph$: a separating four-cycle is exactly a four-cycle in one of these graphs that is not a face of $\graph$. Such a representation may be found in linear time~\cite{Epp-IPL-94}. In an extended graph $E(\graph)$, we allow the external vertices to be included as part of its separating four-cycles.

If $\graph$ is a proper graph with a corner assignment $E(\graph)$, and $C$ is a separating four-cycle in $E(\graph)$ we may form two minors of $\graph$, the \emph{separation components} of $\graph$ with respect to $C$. The \emph{inner separation component} $\graph_C$ is the subgraph induced by the vertices interior to the cycle, and its extended graph $E(\graph_C)$ is the subgraph induced by the vertices on or interior to the cycle, interpreting the vertices of $C$ as a corner assignment for its interior vertices. The
\emph{outer separation component} $E(\graph)\setminus\graph_C$ is formed by replacing the interior of $C$ by a single vertex. We define a \emph{minimal separation component} of $\graph$ to be a minor of $\graph$ formed by repeatedly splitting larger graphs into separation components until no nontrivial separating four-cycles remain. A partition of $E(\graph)$ into minimal separation components may be found in linear time by applying the algorithm for finding all maximal complete bipartite subgraphs $K_{2,i}$ as described above, and then for each such subgraph separating the exterior of the $K_{2,i}$ subgraph from each of the subgraphs within one of the inner faces of the $K_{2,i}$ subgraph.

\begin{lemma}\label{lem:separation}
An extended graph $E(\graph)$ is dual to a one-sided layout if and only if both its inner and outer separation components are dual to one-sided layouts.
\end{lemma}
\begin{proof}
In any layout dual to $E(\graph)$, the region enclosed by the four rectangles of the separating cycle $C$ must be a four-sided polygon, that is, a rectangle. If we modify a one-sided layout of $E(\graph)$ by replacing the contents of this rectangle by a single rectangular area, or by removing the exterior of this rectangle, we obtain one-sided layouts of $E(\graph)\setminus\graph_C$ and $E(\graph_C)$ respectively.

Conversely, suppose we have one-sided layouts of both $E(\graph)\setminus\graph_C$ and $E(\graph_C)$. We may transform the layout of the inner separation component $E(\graph_C)$ so that its bounding box matches the rectangle in the center of $C$ in the layout for the outer separation component $E(\graph)\setminus\graph_C$, and combine these two layouts to obtain a layout of $E(\graph)$. The adjacencies between rectangles and maximal segments of the combined layout are unchanged except for the segments bounding the central rectangle. By the one-sidedness of the layout for the outer separation component, each such segment forms a side of one of the rectangles dual to the vertices of $C$ (the inner rectangle on the other side of the segment has sides that are subsets of the sides of the rectangles dual to $C$), and this property remains true in the combined layout, which is therefore one-sided.
\end{proof}
\begin{cor}
An extended graph $E(\graph)$  is dual to a one-sided layout if and only if all of its minimal separation components are dual to one-sided layouts.
\end{cor}

Thus, if we seek to determine whether an extended graph $E(\graph)$ is dual to a one-sided layout, we may assume without loss of generality that $E(\graph)$ has no nontrivial separating four-cycles. The same idea of cutting the input on separating four-cycles has  been previously applied to the problem of finding sliceable duals for a given proper graph~\cite{DasSur-DAES-01,Mum-PhD-08}.
\begin{figure}[t]
\centering\includegraphics{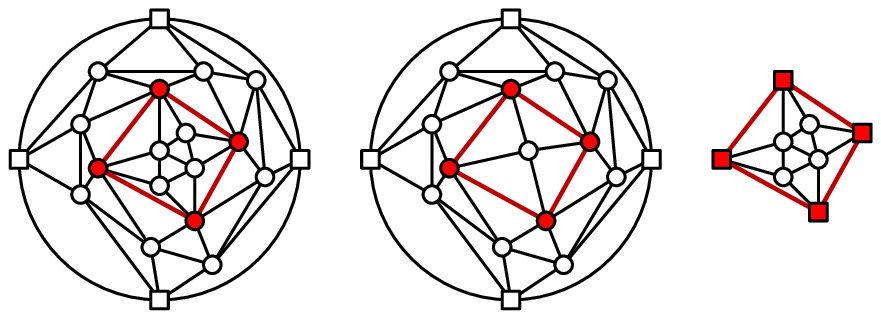}
\caption{An extended graph with a nontrivial separating four-cycle (left), its outer separation component (center), and its inner separation component (right).}
\label{fig:separation}
\end{figure}

\subsection{The distributive lattice of regular edge labelings}

\begin{figure}[t]
\centering\includegraphics{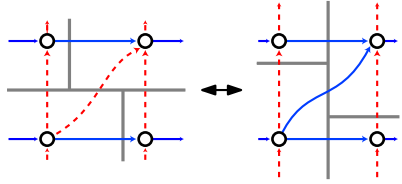}
\caption{A move formed by recoloring the interior of an alternatingly-colored four-cycle in a regular edge labeling, and its effect on the dual rectangular layout. In the case shown, the cycle is not separating: it contains a single edge of $\graph$, but no vertices.}
\label{fig:edge-move}
\end{figure}

Fusy~\cite{Fus-GD-05,Fus-DM-08} (see also \cite{TanChe-ISCAS-90}) defines a family of moves by which one regular edge labeling can be changed to another. Let $C$ be a four-cycle in $E(\graph)$ in which the colors alternate between red and blue around the cycle. Then a move consists of reversing the colors of the edges within $C$; when such a move is made, there can be only one way of setting the orientations of the recolored edges. In a graph with no nontrivial separating four-cycles, each move changes the edge labeling either of a single edge (as shown in Figure~\ref{fig:edge-move}) or of all four edges surrounding a degree-four vertex.  At each of the two or five vertices adjacent to the recolored edges, one of the boundaries between incoming red edges, incoming blue edges, outgoing red edges, and outgoing blue edges shifts by one position in the cyclic ordering of edges around the vertex. These shifts are the same direction for each affected vertex, and can also be interpreted as twisting the boundary between two rectangles in the dual layout by 90 degrees in the opposite direction. Consider the graph with one vertex per regular edge labeling of $E(\graph)$ and with an edge between every two labelings connected by one of these moves; direct each edge of this graph from the labeling in which the boundaries are more clockwise to the labeling in which the boundaries are more counterclockwise. Then this graph of labelings is acyclic and defines a partial ordering on the family of all regular edge labelings of $E(\graph)$. Figure~\ref{fig:8layouts} shows an example, in which the edges in the graph of labelings are directed from the lower labelings to the higher ones.

Moreover, as Fusy shows, the partial order defined in this way is a \emph{distributive lattice}.
A \emph{lattice} is a partially ordered set in which each pair of elements $(a,b)$ has a unique smallest upper bound (such an element is called the \emph{join} of $a$ and $b$ and is denoted $a\vee b$) and a unique largest lower bound (such an element is called the \emph{meet} of $a$ and $b$ and is denoted $a\wedge b$). A \emph{distributive lattice} is a lattice in which the join and the meet operations are distributive over each other: $a\vee(b\wedge c)=(a\vee b)\wedge(a\vee c)$ and $a\wedge(b\vee c)=(a\wedge b)\vee(a\wedge c)$. An element $b$ of a lattice is said to \emph{cover} an element $a$ if $a < b$ and $b$ are immediate neighbors in the lattice, that is, $a < b$ and there exists no element $c$ such that $a < c < b$. In the distributive lattice defined in this way from the regular edge labelings of $E(\graph)$, the covering pairs are exactly the pairs of labelings connected by Fusy's moves. The minimal element of the lattice may be found from any lattice element by repeatedly performing clockwise moves until no more such moves are possible, and the maximal element may similarly be found by repeatedly performing counterclockwise moves. We say that a sequence of moves of the latter type, in which each move is counterclockwise, is \emph{monotone}.

Birkhoff's representation theorem for distributive lattices~\cite{Bir-DMJ-37} states that the elements of any finite distributive lattice may be represented by sets, in such a way that the join and meet operations may be represented by unions and intersections of sets. More precisely, let $P$ be the partial order induced by the subset of the lattice consisting of elements that have exactly one predecessor in the covering relation. Then we may represent any lattice element $x$ by a partition of the partial order into two sets $(L(x),U(x))$ where $L(x)$ consists of the members $y$ of $P$ with $y\le x$ and $U(x)$ consists of the remaining members of $P$. Clearly, $L(x)$ is downward-closed (if $y\le z$ in $P$ and $z\in L(x)$, then $y\in L(x)$) and conversely $U(x)$ is upward-closed. If $x$ and $y$ are two members of the distributive lattice, then $x\le y$ if and only if $L(x)\subset L(y)$ if and only if $U(x)\supset U(y)$, $x\wedge y$ is represented by the partition $(L(x)\cap L(y),U(x)\cup U(y))$, and $x\vee y$ is represented by the partition $(L(x)\cup L(y),U(x)\cap U(y))$. The lattice itself can be reconstructed as the set of all partitions of $P$ into downward- and upward-closed subsets $(L,U)$: each such partition corresponds in this way to a lattice element $x$.

\subsection{The partial order of flippable items}

\begin{figure}[t]
\centering
\includegraphics[width=\textwidth]{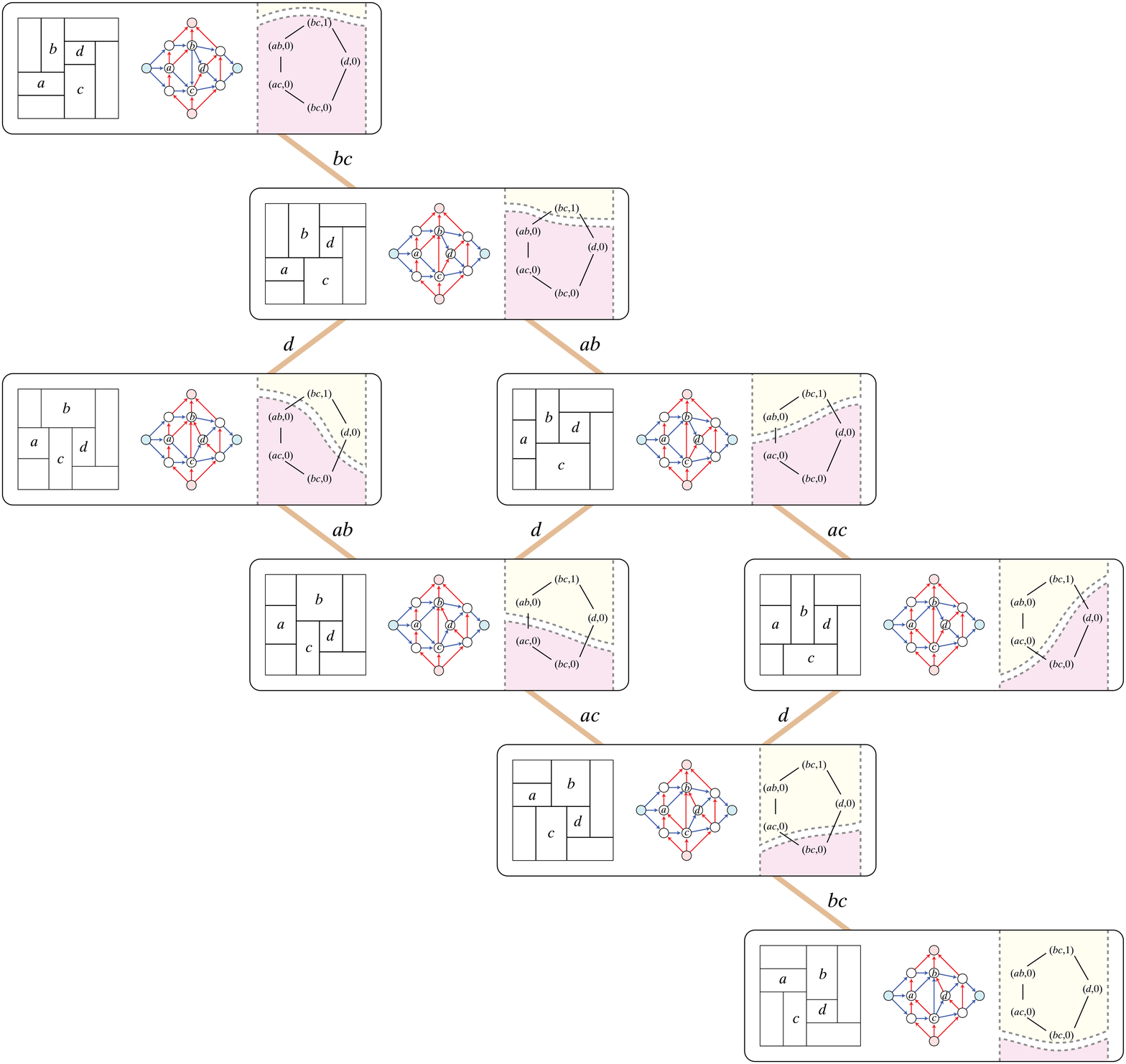}
\caption{The family of rectangular layouts dual to a given extended graph $E(\graph)$, the corresponding regular edge labelings, and the corresponding partial order partitions. Two layouts are shown connected to each other by an edge if they differ by reversing the color within a single alternatingly-colored four-cycle; these moves are labeled by the identity of the edge or vertex contained by the four-cycle.}
\label{fig:8layouts}
\end{figure}
We have seen that the layouts of an extended graph $E(\graph)$ may be described as partitions of a partial order $P$ into a downward-closed and an upward-closed subset; $P$ is the order induced from the distributive lattice of layouts by the subset of layouts that have exactly one downward neighbor.
Our goal in this section is to describe a partial order equivalent to $P$ in a more concrete way, with elements that are not whole layouts themselves but rather that correspond to individual features of rectangular layouts and their dual graphs, in a way that helps us relate the distributive lattice operations more closely to their effect on a layout. Our more concrete partial order, and the partitions of it into subsets $L(\layout)$ and $U(\layout)$ that correspond to each rectangular layout $\layout$, are depicted alongside the layouts in Figure~\ref{fig:8layouts}.

Define a \emph{flippable item} in the extended graph $E(\graph)$ to be either a degree-four vertex $v$ or an edge $e$ that is
not adjacent to a degree-four vertex, with the additional property that there exists some regular edge labeling of $E(\graph)$ in which the four-cycle surrounding $v$ or $e$ is alternately colored and oriented. Thus, a flippable item is the edge that changes color, or the endpoint of a set of four edges that change color, in some move of $E(\graph)$. If $x$ is a flippable item, and $\layout$ is a rectangular layout represented by an element of the distributive lattice of labelings, define $f_x(\layout)$ as the number of moves involving $x$ on any monotone sequence of moves from the minimal lattice element to $\layout$.

\begin{lemma}\label{lem:flipcount}
The number $f_x(\layout)$ is well defined and independent of the monotone path chosen to reach $\layout$ from the minimal element.
\end{lemma}

\begin{figure}
  \hfill
  \begin{minipage}[t]{.44\textwidth}
    \centering
    \includegraphics{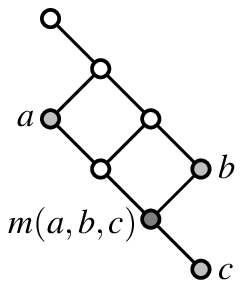}
    \caption{The median of three elements in a distributive lattice.}
    \label{fig:dl-median}
  \end{minipage}
  \hfill
  \begin{minipage}[t]{.44\textwidth}
    \centering
    \includegraphics{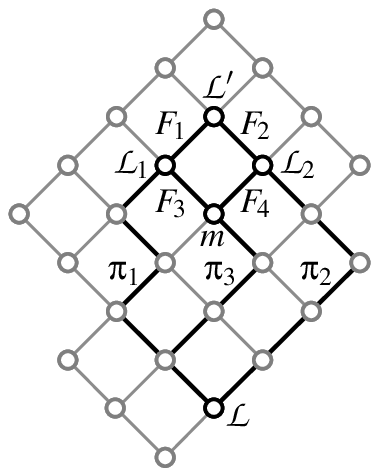}
    \caption{Notation for the proof of Lemma~\ref{lem:flipcount}.}
    \label{fig:lemma5}
  \end{minipage}
  \hfill\hfill
\end{figure}

\begin{proof}
By Birkhoff's theorem, the length of any two upward paths between two
elements of a distributive lattice is equal (it is equal to the size
of the difference of the downward-closed subsets of the partial order
representing those elements). By results of Birkhoff and Kiss~\cite{BirKis-BotAMS-47}, any three elements $a$, $b$, $c$ of a distributive lattice have a unique median
$m(a,b,c) = (a \vee b) \wedge (a \vee c) \wedge (b \vee c) = (a \wedge b) \vee (a \wedge c) \vee (b \wedge c)$
belonging to shortest paths between any two of the three items (Figure~\ref{fig:dl-median}).
We prove by induction the following strengthening of the lemma: let $\layout\le\layout'$ be two layouts. Then for any item $x$, and any two monotone paths from $\layout$ to $\layout'$, $x$ is flipped the same number of times on both paths. Note that the number of times $x$ is flipped, modulo four, must be the same on both paths, as the color and orientation of $x$ may be determined from the number of flips modulo four. As base cases for the strengthening, if the distance from $\layout$ to $\layout'$ is one, there can only be one
monotone path, and if the distance is two, each path can flip $x$ only once while
the number of flips of $x$ on both paths must be the same mod 4, so $x$
must be flipped equally often.

To finish, suppose that we have two monotone paths $\pi_1$ from
$\layout$ to $\layout_1$ and $\pi_2$ from $\layout$ to $\layout_2$, such that we can perform one more upward
flip $F_1$ from $\layout_1$ to $\layout'$ and a flip $F_2$ from $\layout_2$ to $\layout'$. We must show that
the number of flips of $x$ on the two paths $\pi_1F_1$ and $\pi_2F_2$ are equal. Let
$m=m(\layout,\layout_1,\layout_2)$. Then there must exist a path $\pi_3$ from $\layout$ to $m$, and flips
$F_3$ and $F_4$ from $m$ to $\layout_1$ and $\layout_2$ respectively, such that $\pi_3F_3$ and $\pi_3F_4$ are monotone
paths from $\layout$ to $\layout_1$ and $\layout_2$ respectively. By induction the number of
flips of $x$ on $\pi_1$ equals the number of flips of $x$ on $\pi_3F_3$, and the
number of flips of $x$ on $\pi_2$ equals the number of flips of $x$ on $\pi_3F_4$.
Thus, the numbers of flips of $x$ on $\layout_1$ and $\layout_2$ can only differ by one, and the numbers of flips of $x$ on $\pi_1F_1$ and $\pi_2F_2$ can only differ by two. But
again, these numbers of flips must be equal mod 4, so the result
holds.
\end{proof}

\begin{lemma}
\label{lem:flipbound}
The number $f_x(\layout)$ is $O(n)$, where $n$ is the number of rectangles in the layout.
\end{lemma}

\begin{proof}
Define a flipping graph where the nodes are the degree-four vertices and non-degree-four edges of $\graph$ and where two nodes are connected if they belong to the same triangle of $\graph$. In any monotone sequence of moves, whenever a move on $x$ increases $f_x(\layout)$, then $x$ cannot be flipped again until all its neighbors in the flipping graph have been flipped. Therefore, if $x$ and $y$ are adjacent in the flipping graph, $f_x(\layout)$ and $f_y(\layout)$ are always within one of each other. But because the outer edges (or outer degree-four vertices) of the layout can never change color or orientation, the flippable items adjoining them can have $f_x$ at most equal to one. Therefore, the maximum value of $f_x(\layout)$ for any $x$ is at most the length of the shortest path in the flipping graph to one of the boundary nodes, and is $O(n)$.
\end{proof}

Let $\hat\layout$ denote the maximal element in the distributive lattice of labelings.
We define a partial order $P(\graph)$ that has as its elements the pairs $(x,i)$, where $x$ is a flippable element and $i$ is an integer satisfying $0\le i< f_x(\hat\layout)$. Thus, if element $x$ has $k$ different states in different layouts, it participates in $k-1$ pairs of $P$; the pairs correspond not to states but to transitions between states.
In this partial order $P(\graph)$, we define $(x,i)\le (y,j)$ when for all layouts $\layout$ with $f_x(\layout)\le i$, it holds that $f_y(\layout)\le j$; that is, it is not possible to move $f_y$ from $j$ to $j+1$ prior to moving $f_x$ from $i$ to $i+1$. We may represent a layout $\layout$ by the partition of $P(\graph)$ into two subsets $L(\layout)$ and $U(\layout)$, where $(x,i)\in L(\layout)$ when $i < f_x(\layout)$ and $(x,i)\in U(\layout)$ otherwise.

\begin{lemma}
\label{lem:poconstruct}
We can construct $P(\graph)$ from $E(\graph)$ in polynomial time.
\end{lemma}

\begin{proof}
We may compute $f_x(\hat\layout)$ for each $x$, determining the set of elements in $P(\graph)$, by repeatedly performing downward moves in the lattice of layouts until we reach the minimal layout, repeatedly performing upward moves from there until we reach the maximal layout, and counting the number of times a move involves each element~$x$.
The partial order of the pairs $(x,i)$ may be determined from the neighboring objects of $x$ in $E(\graph)$: we may make an upward move involving pair $(x,i)$ in layout $\layout$ if there is no pair $(x,i')$ in $U(\layout)$ with $i'<i$ and when the regular edge labeling corresponding to $\layout$ has the boundaries between incoming red edges, incoming blue edges, outgoing red edges, and outgoing blue edges in a position that would allow such a move at each of the vertices affected by a move at $x$. Each condition that one of these boundaries be in an appropriate position can be characterized by a pair $(x',i')$ that must be moved prior to $(x,i)$ in any monotone sequence of moves starting from the minimal layout, where $x$ and $x'$ are two features of $E(\graph)$ that belong to the same triangle. The minimal pair $(x,i)$ in $U(\layout)$ can be characterized by a constraint that $(x,i') < (x,i)$ in the partial order for each $i'<i$. Thus, by such local considerations, we may find $O(n^2)$ order relations between pairs in $P(\graph)$ that include all covering relations in $P(\graph)$. These order relations define a directed acyclic graph from which the partial order $P(\graph)$ itself may be recovered as the transitive closure.
\end{proof}

In Figure~\ref{fig:8layouts}, each layout is placed next to the corresponding partition of $P(\graph)$ into two subsets $L(\layout)$ and $U(\layout)$. Among the eight layouts in the figure, five of them have exactly one downward neighbor, and these five induce a partial order that is isomorphic to $P(\graph)$. This isomorphism is no coincidence:

\begin{lemma}
$P(\graph)$ is order-isomorphic to the partial order $P$ defined in Birkhoff's representation theorem, and the representation of a layout as a partition of this partial order is the same as the representation in Birkhoff's representation theorem.
\end{lemma}

\begin{proof}
We correspond elements of $P(\graph)$ one-for-one with elements of $P$: each element of $P$ is a layout $\layout$ with only one downward move, to a layout $\layout'$. If this move is on item $x$, then we associate $\layout$ with the pair $(x,i)$ where $i=f_x(\layout)-1=f_x(\layout')$. This pair $(x,i)$ is the single member of the singleton set $L(\layout)\cap U(\layout')$. Conversely, if $(x,i)$ is any pair in $P(\graph)$, we may associate with $(x,i)$ a layout $\layout$ that has only one downward move, as follows: starting from $\hat\layout$, repeatedly perform downward moves that do not reduce $f_x(\layout)$ to $i$ or below, until no more such moves exist; let $\layout$ be the resulting layout.
Each move between two layouts changes both the Birkhoff representation $(L,U)$ and the representation $(L(\layout),U(\layout))$ in corresponding ways. Thus, the two representations are the same. Since $P(\graph)$ and $P$ have a one-to-one correspondence between elements that causes the distributive lattices of their partitions into downward and upward components to have the same elements and the same covering relation, they must be order-isomorphic.
\end{proof}

Thus, we may search through the space of all possible layouts for a given extended graph by instead searching through partitions of $P(\graph)$ into a downward-closed and an upward-closed subset; the possible layouts correspond one-for-one with partitions of this type. The layout represented by a given partition $(L,U)$ may be found by starting from the bottommost layout in the partial order, and repeatedly performing upward moves that do not increase $f_x(\layout)$ (where $x$ is the flippable item involved in the move) to a value $i$ such that $(x,i-1)\in U$, until no more such moves are possible.

\subsection{Order-theoretic characterization of one-sidedness}

We say that a flippable item $x$ is \emph{free} in a layout $\layout$ if there is a move on $x$ available in $\layout$, and \emph{fixed} otherwise. Let $F(\layout)$ denote the set of free flippable items for $\layout$. The following characterization of this set follows immediately from our representation of the distributive lattice of layouts in terms of the partial order $P(\layout)$.

\begin{lemma}
$F(\layout)$ consists of the items $x$ such that some pair $(x,i)$ is a minimal element of $U(\layout)$ or a maximal element of $L(\layout)$.
\end{lemma}

We may then characterize the one-sided layouts in terms of $F(\layout)$:

\begin{lemma}
\label{lem:free1s}
Let layout $\layout$ be dual to an extended graph $E(\graph)$. Then $\layout$ is one-sided if and only if $F(\layout)$ contains no edges of $\graph$.
\end{lemma}

\begin{proof}
If $\layout$ is not one-sided, let $s$ be a maximal segment of $\layout$ with multiple rectangles on both of its sides. Then some edge $e$ of the layout from which $s$ is formed must have as one of its endpoints a T-junction formed by the corners of two rectangles on one side of $s$, and must have on the other endpoint a T-junction formed by the corners of two rectangles on the other side of $s$, as shown in Figure~\ref{fig:edge-move}. These four rectangles form an alternatingly-colored cycle in the regular edge labeling dual to $\layout$, containing a single edge dual to $e$; thus, one may perform a move on this cycle that recolors $e$, as shown in the figure, and $e\in F(\layout)$.
Conversely, if an edge $e$ belongs to $F(\layout)$, the layout edge dual to $e$ must be part of a segment that (because of the alternating coloring of the regular edge labeling cycle surrounding $e$) can be extended in both directions to a maximal segment of $\layout$ that is not one-sided. Thus, in this case, $\layout$ is itself not one-sided.
\end{proof}

Hence, the problem of finding a one-sided layout for $E(\graph)$ becomes equivalent to one of searching for a partition $(L(\layout),U(\layout))$ of the partial order $P(\graph)$ in which the free items consist only of degree-four vertices.

\subsection{Searching for extreme sets}

We have seen in the previous section that one-sided layouts correspond to partitions $(L,U)$ in which the maximal elements of $L$ and the minimal elements of $U$ correspond to degree-four vertices of $\graph$. Each vertex $v$ of $\graph$ can only take one of these roles: it can be a maximal element of $L$ or a minimal element of $U$, but not both, because only one move on $x$ is possible in any layout. Thus, if $\graph$ has $k$ degree-four vertices, then either the maximal elements of $L$ or the minimal elements of $U$ consist of at most $k/2$ members of $P(\graph)$. This motivates the following algorithm for finding one-sided layouts dual to a given graph $\graph$:

\begin{quotation}
\noindent
For each possible extended graph $E(\graph)$ of the given graph $\graph$, and each minimal component $\graph'$ of the extended graph, test whether $\graph'$ has a one-sided layout. If every minimal component has a one-sided layout, form a layout for $E(\graph)$ by gluing these component layouts together. If some minimal component does not have a one-sided layout, then neither does $E(\graph)$.

To test whether $\graph'$ has a one-sided layout, let $k$ be the number of degree-four vertices in $\graph'$, and loop through all sets $S$ consisting of at most $k/2$ members of $P(\graph')$, such that each member of $S$ is a pair $(x,i)$ where $x$ is a degree-four vertex of $\graph$ and all such degree-four vertices are distinct. For each set $S$ of this type, form a partition $(L_1,U_1)$ in which $L_1$ consists of all elements in the partial order that are less than or equal to an element in $S$; if $U_1$ has no minimal elements corresponding to single edges of $\graph$ then return the one-sided layout corresponding to $(L_1,U_1)$. Otherwise, form another partition $(L_2,U_2)$ in which $U_2$ consists of all elements in the partial order that are greater than or equal to an element in $S$. If $L_2$ has no maximal elements corresponding to single edges of $\graph$, return the one-sided layout corresponding to this partition. If neither partition formed in this way from each of the sets $S$ gives rise to a one-sided layout, then $\graph'$ has no one-sided layout.
\end{quotation}

\begin{theorem}
\label{thm:find1s}
Let $K$ be the maximum number of flippable degree-four vertices in any minimal separation component of $\graph$. Then the algorithm described above finds a one-sided layout dual to $\graph$, if one exists, in time $O(n^{K/2+O(1)})$.
\end{theorem}

\begin{proof}
The correctness of the algorithm follows from the sequence of lemmas above. The choice of the extended graph $E(\graph)$ multiplies the number of steps of the algorithm by a factor of $O(n^4)$, and within each minimal component $\graph'$ we loop through $O(n^{K/2})$ sets $S$, performing a polynomial amount of work for each set. Thus, the total time is as stated.
\end{proof}

As special cases, it follows from Lemma~\ref{lem:free1s} that for an extended graph with no flippable degree-four vertex, a one-sided layout exists iff there is exactly one possible layout, for only in that case can $F(\layout)$ be empty. Thus, we may find such a layout by constructing any layout and testing if it is one-sided.
In an extended graph with a single flippable degree-four vertex, a one-sided layout must be either the minimal or the maximal element of the distributive lattice of layouts, for only those two elements can correspond to partitions $(L,U)$ in which $L$ has no maximal elements or $U$ has no minimal elements. Thus, in this case, we need merely construct both layouts and test them for one-sidedness.

\subsection{Fixed-parameter tractability}

Although conceptually straightforward, the algorithm of Theorem~\ref{thm:find1s} is dissatisfactory from the point of view of fixed parameter tractability~\cite{Nie-06}: not just the constant factor in the $O$-notation, but also the exponent of $n$, grows with the parameter $K$. We address this shortcoming by describing an alternative fixed-parameter-tractable algorithm for the same problem.

In a layout $\layout$ of an extended graph $E(\graph)$ with no nontrivial separating four-cycles, define an ordered pair $(v,w)$ of degree-four vertices to be a \emph{stretched pair} if there is no sequence of upward moves from $\layout$ that moves $v$ without moving $w$ and no sequence of downward moves from $\layout$ that moves $w$ without moving $v$. That is, if all relevant pairs belong to the partial order $P(\graph)$,  $(v,f_v(\layout))>(w,f_w(\layout))$ and $(v,f_v(\layout)-1)>(w,f_w(\layout)-1)$. We introduce a special symbol $\emptyset$, and we also define $(v,\emptyset)$ to be a stretched pair if $v$ is in its maximal state ($f_v(\layout) = f_v(\hat\layout)$) and we define $(\emptyset,w)$ to be a stretched pair if $w$ is in its minimal state ($f_w(\layout) = 0$). Thus, the stretched pairs form a directed graph on the vertex set $V$ consisting of the degree-four vertices together with the special symbol $\emptyset$.
We say that a stretched pair $(v,w)$ \emph{fixes} an edge $e$ if $(v,f_v(\layout)-1)\ge (e,f_e(\layout)-1)$ (or $v=\emptyset$ and $f_e(\layout)=0$) and $(w,f_w(\layout))\le (e,f_e(\layout))$ (or $w=\emptyset$ and $f_e(\layout)=f_e(\hat\layout)$).

\begin{lemma}
\label{lem:fixed-is-fixed}
If an edge $e$ is fixed by a stretched pair, $e$ cannot belong to $F(\layout)$.
\end{lemma}

\begin{proof}
Let the stretched pair be $(v, w)$. Because $(v,f_v(\layout)-1) \ge (e,f_e(\layout)-1)$ (or $v=\emptyset$ and $f_e(\layout)=0$), $(e,f_e(\layout)-1)$ is not a maximal element of $L(\layout)$. Similarly, $(e,f_e(\layout))$ is not a minimal element of $U(\layout)$, because $(w,f_w(\layout)) \le (e,f_e(\layout))$ (or $w=\emptyset$ and $f_e(\layout)=f_e(\hat\layout)$). Therefore, $e$ is fixed in $\layout$.
\end{proof}

\begin{lemma}
\label{lem:cyclic-order}
Upward moves on flippable items that are part of the same triangle have a strict cyclical order.
\end{lemma}

\begin{proof}
First assume that the triangle consists of three non-degree-four edges $e_1$, $e_2$, and $e_3$. In every valid regular edge labeling, a triangle (\romannumeral1) cannot be mono-colored and (\romannumeral2) the two edges with the same color must both be oriented towards or from the shared vertex. Let $e_1$ and $e_2$ have the same color in a layout $\layout$. Any move on $e_3$ would violate property (\romannumeral1). Furthermore, it is easy to verify that we cannot do an upward move on both $e_1$ and $e_2$ (if this is allowed by the surrounding edges). Assume that we can do an upward move on $e_1$ resulting in $\layout'$. In $\layout'$ we cannot do a move on $e_2$. Another upward move on $e_1$ can only be performed after performing upward moves on all surrounding edges, including $e_2$ and $e_3$. Hence we can only do an upward move on $e_3$. Continuing this argumentation, the sequence of upward moves on $e_1$, $e_2$ and $e_3$ from $\layout$ must be $e_1, e_3, e_2, e_1, \ldots$. Hence the upward moves on $e_1$, $e_2$ and $e_3$ must follow a strict cyclical order. If a triangle contains a degree-four vertex, only two flippable items $v$ and $e$ are part of this triangle. Using similar argumentation as above, upward moves on $v$ and $e$ have to alternate and hence these moves also must follow a strict cyclical order.
\end{proof}

\begin{lemma}
\label{lem:p-cover}
Let $(y,j)$ cover $(x,i)$ in the partial order $P(\graph)$. Then $x$ and $y$ belong to the same triangle of $\graph$.
\end{lemma}

\begin{proof}
If $(y,j)$ covers $(x,i)$, there must exist a monotone sequence of moves, starting from the minimal element of the distributive lattice of regular edge labelings, such that the penultimate move of the sequence changes $f_x(\layout)$ from $i$ to $i+1$ and the final move of the sequence changes $f_y(\layout)$ from $j$ to $j+1$. But if $x$ and $y$ did not belong to the same triangle of $\graph$, then the four-edge cycle surrounding $y$ would not have its colors or orientation changed by the move on $x$, and the final move on $y$ could have been performed one step earlier, contradicting the assumption that $(x,i)$ is above $(y,j)$ in the partial order.
\end{proof}

\begin{lemma}
\label{lem:mod4}
Suppose $(x,i)$, $(x,i+1)$, $(y,j)$ and $(y,j+1)$ all belong to $P(\graph)$. Then $(x,i)\le (y,j)$ if and only if $(x,i+1)\le (y,j+1)$.
\end{lemma}

\begin{proof}
By Lemma~\ref{lem:p-cover} it suffices to prove that, if $(y,j)$ covers $(x,i)$ then $(y,j+1)\ge (x,i+1)$. For, if we can prove this, then the opposite implication, that if $(y,j+1)$ covers $(x,i+1)$ then $(y,j)\ge (x,i)$ will follow by clockwise-counterclockwise symmetry. And, if $(y,j)\ge (x,i)$ but $(x,i)$ and $(y,j)$ do not form a covering pair, then we can find a chain of covering pairs connecting them in the partial order, and this result will prove that there exists a corresponding chain of order-related pairs four steps higher, proving that $(y,j+1)\ge (x,i+1)$. By Lemma \ref{lem:cyclic-order} and \ref{lem:p-cover}, upward moves on $x$ and $y$ alternate if $(y,j)$ covers $(x,i)$. It easily follows that $(x,i)\le (y,j)$ iff $(x,i+1)\le (y,j+1)$.
\end{proof}

\begin{lemma}
Layout $\layout$ is one-sided if and only if every flippable edge $e$ is fixed by some stretched pair.
\end{lemma}

\begin{proof}
If $e$ is in its minimal state in $\layout$, let $v=\emptyset$; otherwise, $(e,f_e(\layout)-1)$ belongs to $L(\layout)$ and there is a maximal element $(v,f_v(\layout)-1)$ of $L(\layout)$ above it in the partial order. If $e$ is in its maximal state in $\layout$, let $w=\emptyset$; otherwise, $(e,f_e(\layout))$ belongs to $U(\layout)$ and there is a minimal element $(w,f_w(\layout))$ of $U(\layout)$ below it in the partial order. We claim that $(v,w)$ is a stretched pair. For, if all relevant pairs exist in $P(\graph)$, then $P(\graph)$ contains a chain of inequality $(v,f_v(\layout))\ge (e,f_e(\layout))\ge (w,f_w(\layout))$ where the first inequality arises by Lemma~\ref{lem:mod4} and the second comes from the construction of $w$. Using Lemma \ref{lem:mod4}, we also get $(v,f_v(\layout) - 1)\ge (w,f_w(\layout) - 1)$, so $(v, w)$ must be stretched in $\layout$.
\end{proof}

\begin{lemma}
If an edge $e$ is fixed by a stretched pair $(v,w)$ in layout $\layout$, then $e$ is fixed in any layout for which $(v,w)$ are stretched.
\end{lemma}

\begin{proof}
Assume that $(v, w)$ are stretched in $\layout'$, so that $(v, f_v(\layout')) \ge (w, f_w(\layout'))$. This means that $f_v(\layout') - f_v(\layout) = f_w(\layout') - f_w(\layout)$, because if $f_v(\layout') - f_v(\layout) < f_w(\layout') - f_w(\layout)$, then, by Lemma \ref{lem:mod4}, $(v, f_v(\layout) - 1) \ge (w, f_w(\layout))$, which implies that $\layout$ does not exist. Also, because of Lemma \ref{lem:mod4} and $(v, f_v(\layout)) \ge (w, f_w(\layout))$, it must hold that $f_v(\layout') - f_v(\layout) \le f_w(\layout') - f_w(\layout)$. Now let $k = f_v(\layout') - f_v(\layout) = f_w(\layout') - f_w(\layout)$. Because $e$ is fixed in $\layout$, we get that $(e, f_e(\layout) - 1) \le (v, f_v(\layout) - 1)$ and $(w, f_w(\layout)) \le (e, f_e(\layout))$. By Lemma \ref{lem:mod4} we also get that $(e, f_e(\layout) + k - 1) \le (v, f_v(\layout') - 1)$ and $(w, f_w(\layout')) \le (e, f_e(\layout) + k)$. This implies that $k = f_e(\layout') - f_e(\layout)$, from which the lemma follows.
\end{proof}

\begin{lemma}
\label{lem:test-H}
Let $H$ consist of a set of pairs $(v,w)$ that should be stretched. Then in polynomial time we may determine whether there exists a layout $\layout$ of $E(\graph)$ in which all pairs in $H$ are stretched.
\end{lemma}

\begin{proof}
We perform a sequence of upwards moves, starting from the minimal layout, until either a layout satisfying the requirements of $H$ is found or we reach the maximal layout $\hat L$. At each step, if the current layout $\layout$ does not already meet the requirements, it must contain a pair $(v,w)$ that should be stretched but aren't. If $v=\emptyset$, we terminate the search, as no sequence of upward moves can make $w$ minimal if it isn't already. Otherwise, we find a pair $(x,f_x(\layout))$ that is minimal in $U(\layout)$ and below the pair $(v,f_v(\layout))$ (possibly $v=x$), and move upwards on $x$. Such a move must eventually be made to reach any layout that meets the requirements of $\layout$ and is above $\layout$ in the distributive lattice, so each move preserves the set of valid solutions and a solution will eventually be found if one exists.
\end{proof}

\begin{theorem}
Let $K$ be the maximum number of degree-four vertices in any minimal separation component of $E(\graph)$, as before. Then it is possible to find a one-sided layout for $E(\graph)$, if one exists, in time $2^{O(K^2)}n^{O(1)}$.
\end{theorem}

\begin{proof}
As above, we test each minimal separation component separately. Within each minimal separation components, we try all possible choices of the information $H$, consisting of a set of stretched pairs. For each value of $H$, we determine whether the stretched pairs in $H$ fix all of the edges in $E(\graph)$. There are $2^{O(K^2)}$ choices, and each can be tested in polynomial time by Lemma~\ref{lem:test-H}.
\end{proof}

It may be possible to improve the $2^{O(K^2)}$ term in this time bound to $2^{O(K\log K)}$, by using the embedding structure of $E(\graph)$ to restrict the graph of stretched pairs to be a planar graph, but we have not worked out the details of such an improvement.

\section{Layouts with given dual spanning trees}
\label{sec:dualtree}

\begin{figure}[h]
\centering\includegraphics{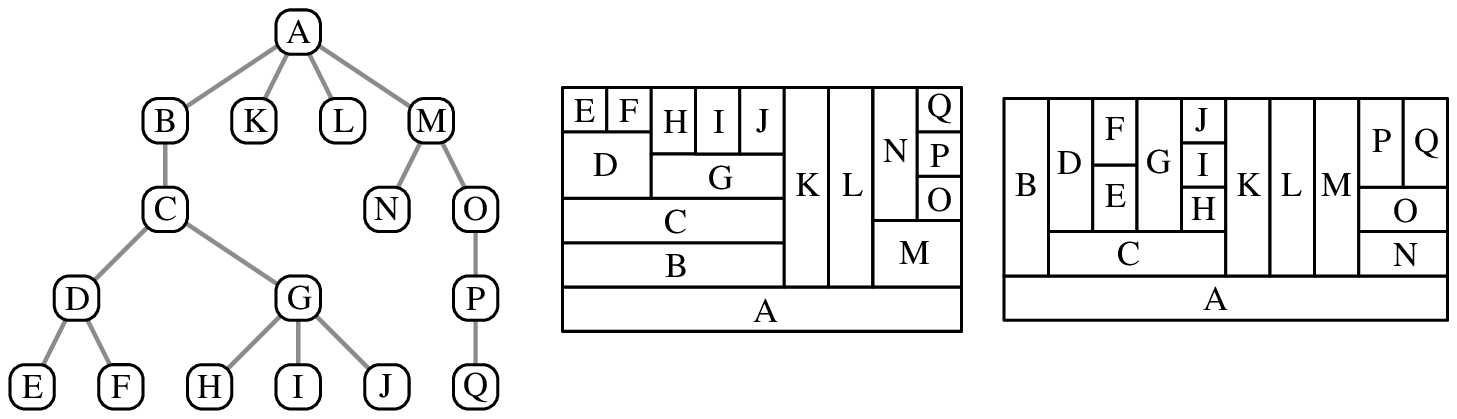}
\vspace{-.25\baselineskip}
\caption{A dual spanning tree $\tree$, Rinsma's non-area-universal layout, and our area-universal layout for~$\tree$.}
\label{fig:rinsma-comparison}
\vspace{-.5\baselineskip}
\end{figure}

Rinsma~\cite{Rinsma1988} considered the question of finding a cartogram for a given weight vector, such that the dual graph $\graph$ has a given tree $\tree$ as its spanning tree.
She showed that, by a simple layout process in which the root of $\tree$ is placed at the bottom of a layout and recursively constructed layouts for its children are placed above it, such a cartogram can always be found. However, her layouts are not, in general, area-universal. For instance, in the layout shown in the center of Figure~\ref{fig:rinsma-comparison}, produced by her algorithm, the line segment with rectangles $D$ and $F$ to its left and with rectangles $G$ and $H$ to the right is not one-sided, showing that the tree in this example leads to a non-area-universal layout according to her algorithm.

However, a simple modification of Rinsma's layout process can be used to generate area-universal layouts that have the given tree as a spanning tree of the dual. The method produces layouts in which the root of a tree either covers the entire bottom edge of the layout or the entire left edge of the layout. For a given tree, to find a layout with the root at the bottom, use the same algorithm recursively to generate layouts for each subtree rooted at a child of the root with the child at the left, and place these subtree layouts in left-to-right order above the bottom root rectangle. Symmetrically, to find a layout with the root on the left, use the same algorithm recursively to generate layouts for each subtree rooted at a child with the child on the bottom, and place these subtree layouts in bottom-to-top order to the right of the root rectangle.
Thus, for a given tree, the layouts with the root at the bottom and with the root at the left are mirror images of each other, as reflected across a line with slope one. The area-universal layout resulting from this algorithm for the same example tree is shown on the right of Figure~\ref{fig:rinsma-comparison}.

\begin{theorem}
For any tree $T$ the algorithm described above finds an area-universal layout, having $T$ as a spanning tree of the dual, in time linear in the size of $T$.
\end{theorem}

\begin{proof}
At each level of the recursion, each child is placed adjacently to the root of its subtree, so $T$ is a spanning tree of the dual, and the algorithm clearly runs in linear time. Each maximal segment of the layout, other than the outer boundaries of the root rectangle, either separates the root of a subtree from its children or one child subtree from the next child subtree. If the segment separates the root of a subtree from its children, it forms a side of the root rectangle, and if it separates one child subtree from the next, it forms a side of the root of the second subtree. Thus, each maximal segment is the side of a rectangle and hence the layout is one-sided. The result follows by Theorem~\ref{thm:3-equivalent-properties}.
\end{proof}

\section{Conclusions and open problems}\label{sec:disc}

We presented a simple necessary and sufficient condition for a rectangular layout to be area-universal. We also described how to find a layout that is equivalent or order-equivalent to a given layout and that realizes a given weight function as a cartogram. Furthermore, we showed how find a one-sided and hence area-universal layout for a given set of adjacency constraints, if such a layout exists. We also investigated similar problems for perimeter in place of area. Unlike much past work on rectangular layouts, we did not restrict our attention to sliceable layouts, dual graphs without separating cycles, or other such special cases.

There remain several questions for further investigation. For instance, our algorithm for finding area-universal rectangular cartograms is not fully polynomial, and it would be of interest to find faster algorithms or determine if it is NP-complete to test whether an area-universal cartogram exists for a given dual graph. If an area-universal cartogram does not exist, but we are given an area assignment or a range of area assignments, can we efficiently find a layout realizing this assignment or assignments? Past work on related problems suggests that such problems might be difficult~\cite{BieGen-TR-05}.

\section*{Acknowledgements}

Work of D. Eppstein was supported in part by NSF grant
0830403 and by the Office of Naval Research under grant
N00014-08-1-1015. B. Speckmann and K. Verbeek are supported by the Netherlands Organisation for Scientific Research (NWO) under project no.~639.022.707.

B. Speckmann would like to thank O. Aichholzer, T. Hackl, and B. Vogtenhuber for stimulating discussions concerning the topic of this paper.

{\small\raggedright
\bibliographystyle{abbrv}
\bibliography{rectilinear}}

\end{document}